

\input vanilla.sty
\tolerance10000
\dimen50=13cm
\advance\dimen50 by -120pt
\divide\dimen50 by 5
\dimen51=\dimen50
\multiply\dimen51 by 2
\advance\dimen51 by 15pt
\dimen52=\dimen50
\multiply\dimen52 by 3
\advance\dimen52 by 30pt
\def\uramii#1#2#3{
\hbox{\vrule\vbox to #2{\hrule\vfil\hbox
to #3{\hfil #1 \hfil}\vfil\hrule}\vrule}}%
\baselineskip=11.5pt
\magnification=\magstep1
\voffset-.7cm
\rm
\nopagenumbers
\pageno=349
\headline={\hss\tenrm\folio\hss}
\null
\hsize 13cm
\vsize 21.4cm
\

\centerline{
{\bf M.Barone} and {\bf F.Selleri}, Editors}

\centerline{
{\bf Advances in Fundamental Physics}}

\centerline{
Hadronic Press, Palm Harbor, Fl U.S.A.}

\centerline{
ISBN $0$-$911767$-$72$-X, Pages $349$-$355, 1995$}

\

\

\

\noindent{\bf RELATIVISTIC DYNAMICS AND SPACE-TIME STRUCTURE}
\vskip.2cm
\noindent{\bf OF FEW-BODY PROCESSES}
\vskip.15cm

\

\

\hskip1.54cm Mladen DAMJANOVI\'C and Zvonko MARI\'C
\vskip.25cm
\hskip1.54cm Institute of Physics, Belgrade, Yugoslavia
\vskip.15cm

\

\

\noindent{\bf 1. INTRODUCTION}

\

The time dependent propagator theory appears to be a suitable
apparatus for the analysis of space-time features of the relativistic
dynamics. We will introduce two kinds of generalizations. The adiabatic
hypothesis is substituted by the disturbative adiabatic hypothesis,
resulting with the corresponding space-time regionalization of scattering
processes. After that we will define a form of the propagator theory for
the description of the few-body processes whose propagation is based on the
family of spacelike surfaces.

The general scheme can be applied to different models of few body
processes also including compositeness levels mixing.

The first part of our investigation includes two broad theoretical
approaches.

One of them is the "traditional" approach for studies of scattering
processes with bound states of the nucleons involved in a process,$^1$ based
on propagator theory$^{2-5}$ with the use of the adiabatic hypothesis,
$^{4,6,1}$ generalized Bethe-Salpeter equation$^{7-8}$ and method of
functional derivation.$^{5,1}$ With appropriate modifications and
generalizations this kind
of approach gives a chance to apply the theory to different classes of
scattering processes as well as for various models by which these processes
are described.

The other theoretical segment is comparative assessment of approaches
for description of few body systems at low and intermediate energies. It
includes many body and few body processes presented at the meson baryon
level, microscopic models (QCD and QCD - inspired models) and hybrid models
as well. Majority of these models are not grounded on the First Principles
of the Quantum Field Theory (QFT). In the section 2. we shall make an
effort to consider a link between these theoretical developments.

The second part of our investigation is based on the fact that in
general forms of the relativistic particle theories, a propagation of a
particle system is formulated on the family of spacelike surfaces in
Minkowski space. For such "one parameter family of spacelike surfaces
filling the whole of space-time, so that one and only one member
$\sigma (x)$ of the family passes through any given point $x$",
introduced by Dyson$^9$ we will
use the name {\it Dyson Family of Surfaces} (DFS).

Instead of using DFS as a mere introductory concept of QFT, it will be
considered as a representative of the general form of relativistic
dynamics.$^{10,12}$ Under such an assumption, the well known forms of
relativistic dynamics: instant, light cone and point ones appear to be the
particular, rigid realizations of an above mentioned general form.

We notice two ways for approaching the relativistic dynamics based on
the DFS. One starts with a time dependent propagator theory for composite
systems. This concept should be "translated" from the traditional language
of the propagation based on the family of instant surfaces to the
presentation of the propagation based on the Dyson family of surfaces.

In this contribution we shall present only main characteristics of our
theoretical approach. The complete form including details of applications
will be published elsewhere.$^{10}$ Some segments of the formalism have been
introduced earlier.$^{11-12}$

\

\

\noindent{\bf 2. EXPANDED TIME-DEPENDENT PROPAGATOR THEORY}

\

The traditional theory which includes the adiabatic hypothesis we
shall call the {\it Standard Time-Dependent Propagator Theory} (STDPT).

Constituents of the process, the particles and their bound states
define initial and final channels:
$$
\{ I_1, ... , I_k\} = \alpha ;\qquad \{ F_1, ... , F_l \} = \beta .\eqno(1)
$$

The amplitudes and propagators are
$$
\chi_\alpha (n)\equiv \ < 0\mid T \bigl( \psi (x_1) ... \psi
(x_n)\bigr)\mid\alpha > ;\quad {\bar\chi}_\alpha (n)\equiv \ <\alpha\mid
T\bigl(\psi (x_1) ... \psi (x_n)\bigr)\mid 0> .\eqno(2)
$$
$$
G_{\alpha\beta}(n; m) = \ <0\mid T\biggl( T\bigl(\psi (x_1) ... \psi
(x_n)\bigr) {\hskip.1cm}T\bigl( \psi (x_1^\prime ) ... \psi
(x_m^\prime )\bigr)\biggr) \mid 0 > .\eqno(3)
$$
Here, $\psi (x_i)$ are renormalized Heisenberg operators.

For (2) and (3) the limiting procedure of Gell-Man and Low$^8$ reads
$$
\bigl[ \psi (x_1) ... \psi (x_n)\bigr]^t=\lim\limits_{x_{io}\to t} T(\psi
(x_1) ... \psi (x_n)) .\eqno(4)
$$
$$
\bigl[ \psi (x_1) ... \psi (x_n)\bigr]^{in, out}=\lim\limits_{t\to
-\infty ,+\infty}\bigl[\psi (x_1) ... \psi (x_n)\bigr]^t .\eqno(5)
$$

Here $\lim\limits_{x_{io}\to t} T$ is notation for the $T$ product in
the limiting instance where two or more times are equated.

One gets the $S$-matrix element $S_{\beta\alpha}={}^{out}
{\hbox{\hskip-.15cm}}<\beta\mid\alpha >^{in}$
by introducing the unit operators
$\sum_\alpha\mid\alpha ><\alpha\mid$ and $\sum_\beta\mid\beta ><\beta\mid$:
$$
\eqalign{
&L(\infty , -\infty ){\hskip.1cm}G_{\alpha\beta} (n; m)= \
<0\mid [\psi (x_1) ... \psi(x_n)]^{out}{\hskip.1cm}[\psi
(x_1^\prime ) ... \psi (x_m^\prime )]^{in}\mid 0 > \ =\cr
&=\sum\limits_{\alpha ,
\beta}\chi_\beta^{out}(n)S_{\beta\alpha}{\bar\chi}_\alpha^{in}(m) .
\cr}\eqno(6)
$$

Above definitions allow the traditional formulation of the {\it Adiabatic
Hypothesis}:$^1$ "No error is introduced into the treatment of the physically
realizable scattering process by a formulation of the theory in which the
interactions among the particles of interest are 'turned off' at remote
times - provided, of course, that the turning-off procedure is sufficiently
gradual that it does not, of itself, create disturbances."

The adiabatic hypothesis introduces a particular regionalization of
the process in the space-time and also introduces a sort of a cluster
decomposition. The propagator (3) can be written in the form
$$
G_{\alpha\beta} (n; m)=G_\beta (n) R_{\alpha\beta}G_\alpha
=\prod\limits_{f=1}^lG_fR_{\alpha\beta}\prod\limits_{i=1}^kG_i.\eqno(7)
$$

The cluster decomposition is seen on the right-hand side of the
expression (7), where $G_i$ and $G_f$ are propagators of the constituents
of the scattering process and $R_{\alpha\beta}$ is a truncated propagator.

If analogous procedure is applied to the amplitudes by writing
$$
{\bar\chi}_\beta^{out}(n)=\prod_{f=1}^l {\bar\chi}_{\beta f} ;\quad
\chi_\alpha^{in}(m)=\prod\limits_{i=1}^k\chi_{\alpha i} .\eqno(8)
$$
and by using (7) and (6), one gets for $S$-matrix the expression
$$
S_{\beta\alpha}={\bar\chi}_\beta^{out}R_{\alpha\beta}\chi_\alpha^{in}
=\prod\limits_{f=1}^l{\bar\chi}_{\beta
f}R_{\alpha\beta}\prod\limits_{i=1}^k\chi_{\alpha i} .\eqno(9)
$$

Space-time content of the adiabatic hypothesis is a segmentation of
the scattering process to the initial, interactional and final phases.
Consequently, in the STDPT, propagator (7) and $S$-matrix (9) are
factorized on the corresponding propagation regions ${\cal R}_i,
{\cal R}_k$ and ${\cal R}_f$ respectively. These regions are separated by two
space-like surfaces.

Our view on the scattering processes will be somewhat different. We
shall retain the supposition that interaction between constituents of
scattering processes could be switched off at the remote times, or
precisely for the incoming and outgoing space-time regions of propagation.
But in addition we shall include a new one by which the mechanism of
inclusion/exclusion is such that it creates the disturbances for
incoming/outgoing channels. We shall suppose also that such effects can be
represented by segments obtained in a factorization of the scattering
picture i.e. that the transition from incoming/outgoing regions of
propagation to the region of "pure" interaction, generally, is not direct
but it goes over the intermediate disturbed phases.

Therefore we formulate the {\it Disturbative Adiabatic Hypothesis} (DAH): In
the framework of the QFT, propagation process can be, generally,
represented by a few-step mechanism. Finite intermediate regions of
disturbance which can be represented by a superposition of the physically
realizable configurations for the corresponding channels are compatibly
connected with the central, interaction region. Interaction among the
constituents of the scattering process is "turned-off" for the incoming and
outgoing propagation regions, so that initial and final configurations are
corresponding to them.

The motivations for the above reformulation are mainly the following.

Firstly, they are of the conceptual nature. The intermediate phases
have been tacitly already introduced (by using, for example, off mass shell
contributions) and different cluster decompositions are often introduced ad
hoc without explication of whether it is consistent or not with the general
procedure.

Secondly, in applications of the standard formulation one often finds
the incompatibility of the interaction effects originating from scattering
amplitudes and from truncated propagators. That can be easily seen in the
analysis of the $\gamma D \to pn$ process.$^{13}$

Thirdly, the idea was to find consistent and in the same time flexible
formulation which could be applied to many concrete scattering situations
where these ones appear as particular cases.

In the {\it Expanded Time-Dependent Propagator Theory} space-time content
will be correspondingly reacher. Two additional domains of scattering
process appears. Disturbed channels of the initial and final configurations
correspond to the incoming and outgoing disturbation regions ${\cal
R}_{id}$ i.e. ${\cal R}_{fd}$. So one can visualize the sequence of the
five regions: ${\cal R}_i, {\cal R}_{id}, {\cal R}_k, {\cal R}_{fd}$ and
${\cal R}_f$ separated by the four space-like surfaces. DAH causes the
corresponding factorization of the representatives of the scattering theory.

In our formulation the propagator for the process $\alpha\to\beta$ is written
$$
G_{\alpha\beta}=G_\beta{\bar
K}_\beta^dR_{\alpha\beta}K_\alpha^dG_\alpha .\eqno(10)
$$

Here ${\bar K}_\beta^d$ and $K_\alpha^d$ correspond to the effect of
disturbation introduced.

With them the generalized expanded interaction term $M_{\alpha\beta}$ reads
$$
M_{\alpha\beta}={\bar K}_\beta^dR_{\alpha\beta}K_\alpha^d .\eqno(11)
$$

The $S$-matrix elements gets the form
$$
S_{\beta\alpha}={\bar\chi}_\beta^{out}{\bar
K}_\beta^dR_{\alpha\beta}K_\alpha^d\chi_\alpha^{in} ,\eqno(12)
$$
and the new objects, "disturbation amplitudes" read
$$
\chi_\alpha^d\equiv K_\alpha^d\chi_\alpha^{in} ;\quad{\bar\chi}_\beta^d
\equiv{\bar\chi}_\beta^{out}{\bar K}_\beta^d .\eqno(13)
$$

The process defined by DAH is visualized by the following scheme

$$
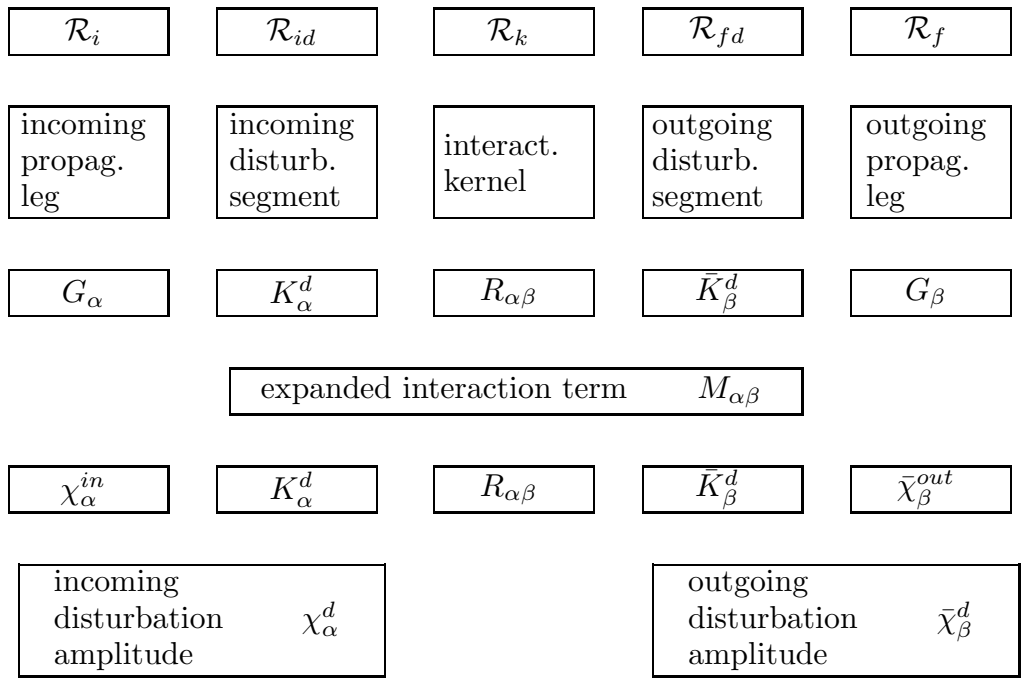
\vbox{\halign{\tabskip=15pt#&#&#&#&#\cr
\uramii{\hbox{${\cal R}_i$}}{1.3\baselineskip}{\dimen50
}&\uramii{\hbox{${\cal R
}_{id}$}}{1.3\baselineskip}{\dimen50}&\uramii{\hbox{${\cal R
}_k$}}{1.3\baselineskip}{\dimen50}&\uramii{\hbox{${\cal R
}_{fd}$}}{1.3\baselineskip}{\dimen50}&\uramii{\hbox{${\cal R
}_f$}}{1.3\baselineskip}{\dimen50}\cr\noalign{\vskip 15pt}
\uramii{\vbox{\hbox{incoming}\hbox{propag.}\hbox{leg
}}}{3.1\baselineskip}{\dimen50}&\uramii{\vbox{\hbox{incoming}\hbox{disturb.
}\hbox{segment}}}{3.1\baselineskip}{\dimen50}&\uramii{\vbox{\hbox{interact.
}\hbox{kernel}}}{3.1\baselineskip}{\dimen50}&\uramii{\vbox{\hbox{outgoing
}\hbox{disturb.}\hbox{segment}}}{3.1\baselineskip}
{\dimen50}&\uramii{\vbox{\hbox{outgoing}\hbox{propag.}\hbox{leg
}}}{3.1\baselineskip}
{\dimen50}\cr\noalign{\vskip 15pt}\uramii{\hbox{$G_\alpha$}}{1.3\baselineskip}
{\dimen50}&\uramii{\hbox{$K_\alpha^d$}}{1.3\baselineskip}
{\dimen50}&\uramii{\hbox{$R_{\alpha\beta}$}}{1.3\baselineskip}
{\dimen50}&\uramii{\hbox{${\bar K}_\beta^d$}}{1.3\baselineskip}
{\dimen50}&\uramii{\hbox{$G_\beta$}}{1.3\baselineskip}
{\dimen50}\cr\noalign{\vskip 15pt}
&\multispan3\uramii{\hbox{expanded interaction term
}\hfil\hfil$M_{\alpha\beta}$}{1.3\baselineskip}
{\dimen52}&\cr\noalign{\vskip 15pt}
\uramii{\hbox{$\chi_\alpha^{in}$}}{1.3\baselineskip}
{\dimen50}&\uramii{\hbox{$K_\alpha^d$}}{1.3\baselineskip}
{\dimen50}&\uramii{\hbox{$R_{\alpha\beta}$}}{1.3\baselineskip}
{\dimen50}&\uramii{\hbox{${\bar K}_\beta^d$}}{1.3\baselineskip}
{\dimen50}&\uramii{\hbox{${\bar\chi}_\beta^{out}$}}{1.3\baselineskip}
{\dimen50}\cr\noalign{\vskip 15pt}
\multispan2\uramii{$\vcenter{\hbox{incoming}\hbox{disturbation
}\hbox{amplitude}}$\hfil\hfil$\chi_\alpha^d$}{3.1\baselineskip}
{\dimen51}& &\multispan2\uramii{$\vcenter{\hbox{outgoing}\hbox{disturbation
}\hbox{amplitude}}$\hfil\hfil${\bar\chi}_\beta^d$}{3.1\baselineskip}
{\dimen51}\cr
}}
$$

\noindent First line corresponds to the domains of the process, second to the
propagational and interactional content of the segments, third to the
propagator and fifth to the $S$-matrix. Forth and sixth line illustrate the
joining of disturbation effects.

The contact with experimental situation is model dependent. The
interaction is given with the form of the expanded interaction term
$M_{\alpha\beta}$. In addition it is necessary to define the disturbation
effects. The remote time amplitudes $\chi_\alpha^{in}$ and
${\bar\chi}_\beta^{out}$ are determined phenomenologically.

The general formalism contains also two procedures which are
independent and mutually compatible.

One of them consist in trying to find "natural" representation of
disturbation effects in the form
$$
K^d=G^dV ;\quad {\bar K}^d={\bar V}G^d .\eqno(14)
$$

Propagators $G^d$ are related to the finite domains. They have the same
physical meaning as propagational segments of a truncated propagator
$R_{\alpha\beta}$ which appear in the course of the functional
derivation.

The second procedure consists in the series expansion
of $M_{\alpha\beta}$ and reads
$$
M_{\alpha\beta}=\bigl( {\bar K}_\beta^{d(0)}+{\bar
K}_\beta^{d(1)}+...\bigr)\bigl(
R_{\alpha\beta}^{(0)}+R_{\alpha\beta}^{(1)}+...\bigr)\bigl(
K_\alpha^{d(0)}+K_\alpha^{d(1)}+...\bigr) .\eqno(15)
$$
This expansion is not perturbative series but it reflects the content of
Bethe-Salpeter equation with the precise meaning of spatio-temporal
composition of interaction effects.

One needs an additional analysis of terms in order to achieve a
compatibility of the interaction contributions independently of their
origin.

\

\

\noindent{\bf 3. PROPAGATION ON THE DYSON FAMILY OF SURFACES}

\

As a second aspect of a time dependent presentation of the scattering
processes we shall consider (I) the covariant generalization of the
space-time propagation of the constituents of a scattering process, and
(II) the analysis of the causality conditions which appear in connection
with the mentioned generalization.

\noindent (I)   One uses a one parameter DFS $f^a(\sigma )$. For the
members of the family of manifolds in Minkowski space the inclusion relation
$\sigma < \sigma_o$ is defined. This allows us to generalize the time step
function by defining chronological step function $\Theta_\sigma$ which
appears as a distribution in Minkowski space.
Analogously, generalized chronological product $T_\sigma$ is introduced.

By the above constructs one defines in a general manner the propagator
$G$ and the amplitudes $\chi$ and ${\bar\chi}$ for a particular channel, i.e.
$$
G_{\alpha\beta}(n ; m)= \ <0\mid T_\sigma\biggl( T_\sigma \bigl(\psi (x_1)
... \psi (x_n)\bigr) {\hskip.1cm}T_\sigma \bigl(\psi (x_1^\prime )
... \psi (x_m^\prime)\bigr)\biggr)\mid 0> ,\eqno(16)
$$
$$
\chi_\alpha (n)= \ < 0\mid T_\sigma \bigl(\psi (x_1) ... \psi (x_n)\bigr
)\mid\alpha
> ; \quad {\bar\chi}_\alpha (n)= \ <\alpha\mid T_\sigma \bigl(\psi (x_1) ...
\psi (x_n)\bigr)\mid 0> .\eqno(17)
$$
\vskip.2cm
\noindent which are covariant generalizations of the usual ones.

The above presentation is realized in such a manner that allows us to
perform time dependent factorization of the scattering processes.

\noindent (II) The implicit correspondence between relevant families of the
space-like surfaces and specific scattering processes is a distant
reflection of the connection between geometry and dynamics which is
practically impossible to study by a standard techniques of the QFT.

Additional results could be achieved by considering the causality
conditions. For the propagator of scattering processes only one condition is
imposed by temporal $T$ or chronological $T_\sigma$ product. Thregulates the
succession of the events. But a propagator contains non causal

contributions as well. It includes time-like separation of the constituents
going to interact and arbitrarily large space-like separation of the basic
constituents of the bound states. This poses the well known problem of
scattering theory of composed systems.

If we take into account nonlocality of constituents we can consider
causal propagation by introducing a specific cut-off.

Mathematically, one can introduce the class of the families of
space-like surfaces $\Sigma ^A(\sigma ) = \cup_{a\in A}f^a(\sigma
)$ for which one defines the characteristic function $P_{\Sigma ^A}$.
We shall impose that the parameterization of
families which form a class is continuous and finite. If one defines the
corresponding space-like domains $D^A=(\sigma\in f^a(\sigma ) \ \mid \
a\in A)$ then to the family of the continuously deformable domains
$\Sigma (D^A)$ corresponds the class of the families of
space-like surfaces $\Sigma^A (\sigma )$.

The propagators defined through the corresponding characteristic
functions remains functionals whose arguments lie in the whole space $M^n$.
However, the significant contributions are only those coming from the
causal separation of their arguments.

The fact which we would like to describe quantitatively is the
following: in the course of the scattering process its constituents, the
particles and the bound states, are in the corresponding space-time domains.
So, it is natural to demand that propagators and amplitudes reflect mutual
correlations and collective effects coming from the interaction.

In addition, we expect compatibility between the disturbance mechanism
and the outlined one. As a consequence all the recoil and retardation
effects which have unpleasant feature in the standard formulation appear
now in the common footing with the whole dynamics.

\

\

\noindent{\bf 4. APPLICATIONS}

\

An example is the coherent pion photoproduction on deuteron:
$\gamma d\to \pi d$. By using (12), (11) and
development (15) and by keeping the zeroth and first order of the meson
exchange terms, the $S$-matrix (12) gets the form:
$$
S_{\beta\alpha}={\bar\chi}_\beta^{out}\bigl( {\bar
K}_\beta^{d(0)}R_{\alpha\beta}^{(0)}K_\alpha^{d(0)}+{\bar
K}_\beta^{d(0)}R_{\alpha\beta}^{(1)}K_\alpha^{d(0)}+{\bar K}_\beta^{d(0)}
R_{\alpha\beta}^{(0)}K_\alpha^{d(1)}+{\bar
K}_\beta^{d(1)}R_{\alpha\beta}^{(0)}K_\alpha^{d(0)}\bigr)\chi_\alpha^{in}
.\eqno(18)
$$
\vskip.2cm
As a result of the application of the DAH all first order terms are
mutually compatible and consequently it is possible to make their
consistent ressumation. We note that it is not the case in the standard
procedure (not using DAH), for example for $\gamma d\to pn$ process.$^{13}$

The main result is the new structure of the transition operator expressed
through the one-nucleonic variables.

For a given energy region one has to specify the disturbation
input, in particular the interaction kernel and the type of meson-nucleon
coupling.

If one considers the energy region of the first few pion nucleon
resonances one might choose that exchanged meson is a pion and the
coupling is a pseudoscalar one. Inclusion of these assumptions leads to the
model studied.$^{14,15}$ The structure of $S$-matrix appears to be a
superposition of the impulse approximation, rescattering of pion on
nucleons, and double pion photoproduction on nucleon with subsequent
absorption of one pion by other nucleon.

The above exposed theoretical scheme is so general that it allows its
application to reactions at QCD level as well as for hybrid models. Details
will be published elsewhere.

\

\

\noindent{\bf REFERENCES}

\

\item{1)}{A. Klein, C. Zemach, {\it Phys. Rev.} {\bf 108}: 126 (1957).}
\item{2)}{R.P. Feynman, {\it Phys. Rev.} {\bf 76}: 749,769 (1949).}
\item{3)}{J. Schwinger, {\it Phys. Rev.} {\bf 82}: 914 (1951);
Proc. N. A. S. (USA) {\bf 37}: 452 (1951).}
\item{4)}{K. Nishijima, {\it Prog. Theor. Phys.} {\bf 10}: 549 (1953);}
\item\item{{\it Prog. Theor. Phys.} {\bf 12}: 279 (1954); {\it Phys. Rev.}
{\bf 111}: 995 (1958).}
\item{5)}{H. Umezawa, A. Viskonti, Nuovo Cim. {\bf 1}: 1079 (1955).}
\item{6)}{K. Nishijima, {\it Prog. Theor. Phys.} {\bf 17}: 765 (1957).}
\item{7)}{E.E. Salpeter, H.A. Bethe, {\it Phys. Rev.} {\bf 84}: 1232 (1951).}
\item{8)}{M. Gell-Mann, F.E. Low, {\it Phys. Rev.} {\bf 84}: 350 (1951).}
\item{9)}{F.J. Dyson, {\it Phys. Rev.} {\bf 75}: 486 (1948).}
\item{10)}{M. Damjanovi\'c, Z. Mari\'c, In the course of preparation.}
\item{11)}{M. Damjanovi\'c, Z. Mari\'c, {\it Proc. of the Eight Meeting of
Yugoslav Nuclear and Particle Physicist}, Portoro\v z, Yug.: 69 (1985).}
\item{12)}{M. Damjanovi\'c, Z. Mari\'c, XIth European Conference on
Few-Body Physics, Fontevraud, France, {\it Abs. of contr. papers}:
21 (1987); {\it Ibid} 20 (1987).}
\item{13)}{L.D. Pearlstein, A. Klein, {\it Phys. Rev.} {\bf 118}:
193 (1960).}
\item{14)}{C. Lazard, Z. Mari\'c, Dj. \v Zivanovi\'c, {\it J. Phys.}
{\bf G5}: 1549 (1979).}
\item{15)}{C. Lazard, Z. Mari\'c, {\it J. Phys.} {\bf G9}: L59 (1983).}
\end